\documentclass[aps,prb,twocolumn,floatfix,footinbib,showpacs,superscriptaddress]{revtex4-1}
\usepackage{graphicx}
\usepackage{amsfonts,amsmath,amssymb}
\usepackage{amsthm}
\usepackage{dsfont,bm}
\usepackage{color}
\usepackage{soul} %Include when compairing comments
\usepackage{amsbsy}
\usepackage[colorlinks=true,linkcolor=blue,pagecolor=blue,filecolor=blue,menucolor=blue,urlcolor=blue,citecolor=blue,anchorcolor=blue]{hyperref}%
\usepackage{multirow}
\usepackage{mathrsfs}

\usepackage{amsbsy}
\usepackage{times} 
\usepackage{dcolumn}

\newcommand*{\rttensor}[1]{\overline{\overline{#1}}}

\usepackage{lineno}
\begin{document}

\title{Controllable Nonreciprocal Optical Response and Handedness-Switching in Magnetized Spin-Orbit Coupled Graphene}

\author{Mohammad Alidoust}
\affiliation{Department of Physics, Norwegian University of Science and Technology, N-7491 Trondheim, Norway}
\author{Klaus Halterman}
\affiliation{Michelson Lab, Physics Division, Naval Air Warfare Center, China Lake, California 93555, USA}

%==========
\begin{abstract}
 Starting from a low-energy effective Hamiltonian model, 
 we theoretically calculate the dynamical optical conductivity and 
 permittivity tensor of a magnetized graphene layer with Rashba spin-orbit coupling (SOC). 
 Our results reveal a transverse Hall conductivity correlated with the nonreciprocal longitudinal conductivity. 
 Further analysis illustrates that for   intermediate
  magnetization strengths, the 
   relative magnitudes of
 the  magnetization and 
 SOC can be  identified experimentally by
   two 
 well-separated peaks in the 
 dynamical optical response (both the longitudinal and transverse components) as a function of photon frequency. 
 Moreover, the frequency-dependent
 permittivity tensor  is obtained for a wide range  of 
 chemical potentials and magnetization strengths.
  Employing experimentally realistic  parameter values, 
 we calculate the circular dichroism of a representative 
 device consisting of  magnetized spin-orbit coupled graphene and a dielectric insulator layer, backed by a 
 metallic plate. The results reveal
  that this device has different relative absorptivities for  
 right-handed and left-handed 
  circularly 
  polarized electromagnetic waves.
    It is found that the magnetized spin-orbit coupled graphene 
    supports strong handedness-switchings, 
    effectively controlled by varying the chemical potential and magnetization strength with respect to the SOC strength.   
\end{abstract}

\date{\today}

\maketitle
%=========

\section{Introduction}

%% graphene 

Graphene is a two-dimensional (2D) planar honeycomb arrangement of carbon atoms
with one-atom thickness. 
The isolation of a graphene sheet was first officially reported in 2004.\cite{graphene} This 
groundbreaking feat
has inspired the isolation 
and creation of several novel  single-layer materials 
that are one-atom thick, such as silicene and phosphorene,
and has fueled the growth of a larger family of 2D materials.\cite{silicene,phosphorene} 
The most interesting electronics characteristics of graphene 
are accessible within a narrow energy window 
close to
the Fermi level.\cite{RMP-2009-Neto} The dispersion of quasiparticles near the
Fermi level follows a linear relation as a function of momentum. 
This  property facilitates the study of relativistic Dirac fermions within a
 practical  platform. \cite{RMP-2008-Beenakker} Additionally, graphene can withstand  relatively high strains
  without rupturing, and supports
a   tunable chemical potential. These intriguing discoveries and advancements have
made graphene attractive for both fundamental science and
 next-generation devices and technologies. \cite{RMP-2009-Neto,G.G.Naumis,S.Sarma}  

On its own, graphene has 
a
negligible band gap and 
limited spin-related features such as magnetization and spin-orbit coupling (SOC) \cite{soc_intrinsic}. 
Due to this,  the use of  free standing
graphene in practical devices is 
still limited. To make graphene more suitable for 
technology-oriented applications and explore interesting fundamental phenomena, 
it is important to capitalize on additional effects such as 
 the interplay of Dirac fermions with superconductivity, magnetization, and SOC.
  Along these lines,  experimentally feasible approaches \cite{exp_f1,ex2,ex3}
  involve  the exploitation of proximity effects,
  whereby  the
  magnetism and SOC can be extrinsically\cite{rashba_gr2} induced into graphene by close
  contact with other materials \cite{D.Marchenko,J.Balakrishnan,A.Avsar,prox4,Z.Wang}. The proximity-induced magnetism and SOC in graphene is more appropriate than chemical doping as the former approach preserves the chemical properties of graphene and the 
  quality of the
  graphene lattice remains nearly intact. \cite{D.Marchenko,J.Balakrishnan,A.Avsar,prox4,Z.Wang} 
   This idea has driven numerous efforts
   both theoretically and experimentally to shed light on various aspects of superconducting \cite{gsuper,halt2}
    magnetized \cite{mag_gr2,mag_gr1,rashba_gr}, or spin-orbit coupled graphene \cite{rashba_gr,equalspin3,equalspin4,soc_gr3,soc_gr4}. 
    
   On both the
   experimental and theoretical  fronts,  transport measurements
    have found excellent agreement with the low-energy effective models \cite{RMP-2009-Neto,S.Sarma,RMP-2008-Beenakker}. 
    Indeed, there have been several experiments carried out in this direction so far,
     reporting successful proximity-induced phenomena in graphene \cite{exp_f1,ex2,ex3}.
     Nonetheless, there is no clear-cut evidence and estimation of the type and strength of the induced SOC in graphene. 
     For example, in Refs.~\onlinecite{exp_f1,ex2,ex3}, 
     ferromagnetism was induced in
     graphene by placing it in close contact with yttrium iron garnet (YIG), which is  a magnetic insulator with SOC. 
      The corresponding  signatures of
      the induced SOC and magnetization into graphene were observed 
      through transverse Hall current measurements;
      however, a clear-cut
      picture  of the type  and strength of the
      SOC remains elusive.         

%% optics
When a plane circularly polarized electromagnetic (EM) wave
interacts with some materials, there can be differences in the 
 absorption of  left-handed (LH) and right-handed (RH)
 circularly polarized light.
 When the absorptance 
 of the incident
 circularly polarized beam depends on the handedness,
 the material possesses optical circular dichroism (CD) \cite{circD}.
Moreover, when tailor-made material platforms support ``handedness-switching''
 under external controls, 
 the system can change whether it predominately absorbs RH or LH
 polarized waves. The corresponding absorptance signatures can 
reveal valuable spin information and, possibly, intrinsic quantum details of the system \cite{circ1,circ2,circ3}.
 Such material platforms with nonzero intrinsic CD can be  
  used in nanodevices as a circular polarization filter
with
 atomic-scale control,
 and
 programmable optoelectronic devices with high-speed switching.

 Having a low-power atomistic-scale mechanism to control electric-field rotation and absorption of EM waves is a desirable capability for modern integrated and compact optical nanodevices, sensors, and detectors \cite{I.K.Reddy,L.D.Barron}. 
 Control of the polarization, phase, and magnitude of reflected and transmitted EM waves has been explored using 
 metamaterials  \cite{ws1,ws2,V.C.Nguyen,S.Feng,K.Halterman,B.-X.Wang,X.Wu,K.Sonowal,Q.Chen,R.Sabri,A.Forouzmand}
 and patterned metasurfaces \cite{C.-C.Chang,M.Barkabian,M.Zare,M.Nickpay,P.Rezaei,P.Zamzam,H.Zhang,L.Hu,F.Li,M.Amin}. 
 Nevertheless, fabrication challenges in creating metasurfaces, in addition to the low efficiency and limited compactness of the final structure, limit the effectiveness of this approach.  Moreover, precise control over the dynamical and widely modulating system 
 parameters is needed \cite{C.-C.Chang,M.Barkabian,M.Zare,M.Nickpay,P.Rezaei,P.Zamzam,H.Zhang,L.Hu,F.Li,M.Amin}, 
 thus 
 making atomic-scale optoelectronic elements with  controllable 
 material properties a more favorable alternative \cite{A.Y.Zhu}.

\begin{figure}[t!]
\includegraphics[width=0.45\textwidth]{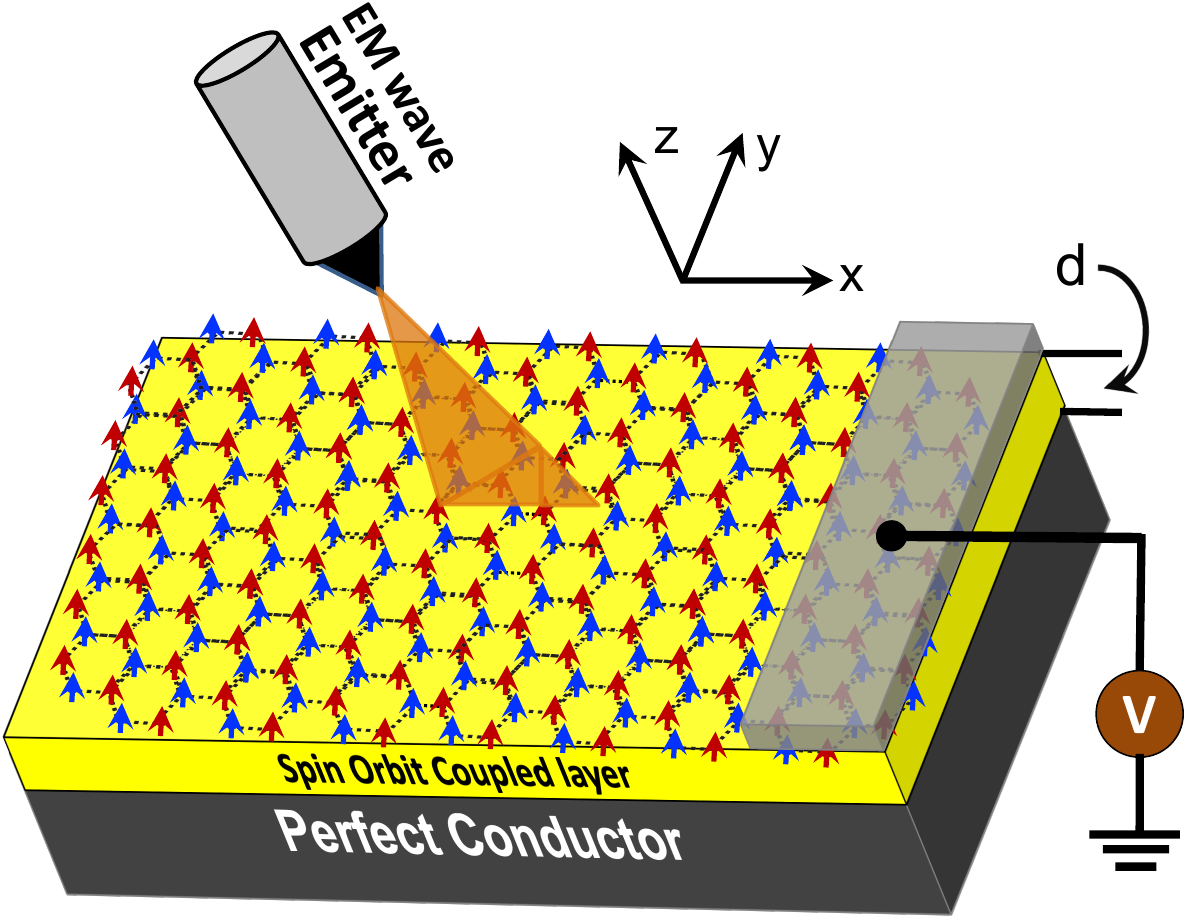}
\caption{\label{fig1} (Color online). The representative 
setup for revealing handedness-switching predicted in this paper. 
The graphene layer is deposited on top of an insulator layer with strong SOC and a thickness of $d$. 
A perfectly conducting back plate accompanies the system to generate strong reflection of 
the electromagnetic wave, which is incident upon the graphene side of the structure. 
The reflected wave is then collected by a detector. We assume that the 
graphene sheet
 is located in the $xy$ plane. 
 The small arrows show the orientation 
 of magnetization on the carbon sites. The blue and red regions indicate the A and B sublattices of graphene, respectively. The carrier density and chemical potential can be controlled by a gate voltage, $V$.}
\end{figure}

In this paper, we have considered a graphene layer with extrinsically induced Rashba SOC and magnetization. 
Starting from a
microscopic Hamiltonian and employing a Green's function approach, 
we derive expressions for the components of 
the dynamical optical conductivity tensor, and thereby
obtain the associated frequency-dependent permittivity tensor. 
Our results illustrate that the
 optical conductivity acquires a strong peak associated 
 with SOC at frequencies close to the 
 corresponding SOC
 energy.
  A nonzero magnetization of moderate strength perpendicular to 
 the
 graphene plane results in a clear second peak in the frequency-dependent optical conductivity, well separated from the SOC peak. To confirm our findings, we analyze their origins through visualizing various possible interband and intraband optical transitions in the band structure. Therefore, we find that the optical conductivity and dielectric response 
 can be practical tools for identifying the presence of SOC induced in graphene,
 and estimating its strength unambiguously. Moreover, upon calculating
 the  CD, we demonstrate that a nanoscale device of the type schematically shown in Fig.~\ref{fig1} 
 can  efficiently control handedness-switching by tuning
 the chemical potential and magnetization strength.  
 Our results and findings may serve as a possible scenario for the physical origins
 of a nonzero CD observed in
   a recent experiment involving a graphene system deposited on a $\rm SiC$ substrate \cite{C.Hwang}.
   This would suggest
   that the interaction of graphene and a substrate can result in SOC and a
   nonzero magnetism.     

% organization

The paper is organized as follows. In Sec.~\ref{sec:theory},
 the theoretical model is outlined, including the model Hamiltonian. In Sec.~\ref{sec:optcond}, the analytical and numerical 
 evaluations of the dynamical optical conductivity are performed. In Sec.~\ref{sec:hand}, the finite-frequency dielectric tensor is discussed 
 and the response of the magnetized spin-orbit coupled graphene to polarized EM waves is analyzed. 
 Finally, we summarize the results  with concluding remarks in Sec.~\ref{sec:concl}.

\section{Formulation and framework}\label{sec:theory}

Graphene atoms are bonded through the $p$-orbital electrons. 
The electronics 
consequences of these interactions for
 low energies close to the
 Fermi level can be properly described by a tight-binding model.\cite{RMP-2009-Neto} 
 The tight-binding model can be further simplified into an effective Hamiltonian model 
 around 
 the Fermi level without missing any important physics \cite{RMP-2009-Neto}.
 Unlike the more complicated computational methods, the effective Hamiltonian model provides 
 more clarity  into the fundamental physics of systems. 
 Also, it has long been proven 
 that by contrasting results to experimental observations, the effective Hamiltonian captures the most interesting physics in graphene 
  at low energies. \cite{RMP-2009-Neto,S.Sarma,RMP-2008-Beenakker} 
  Both magnetization and SOC can be extrinsically induced into graphene by 
  virtue of the proximity effect involving the appropriate materials. \cite{exp_f1,ex2,ex3} 
  The effective Hamiltonian of graphene with SOC and magnetization can be expressed as \cite{soc_gr1,soc_gr2},
\begin{subequations}\label{hamil}
\begin{align}
&\mathcal{H}{=}\int d \textbf{k}\Psi^\dagger(\textbf{k}) H(\textbf{k})\Psi(\textbf{k}),\\
&H(\textbf{k}){=}\hbar v_\text{F} \textbf{k}\cdot\boldsymbol{\tau}+\alpha (\boldsymbol{\sigma}\times\boldsymbol{\tau})_z  +\textbf{h}\cdot \boldsymbol{\sigma}-\mu.
\end{align}
\end{subequations}
The particles moving within the plane of graphene have momentum $\textbf{k}=(k_x,k_y)$. The particles' velocity at 
the
Fermi level is approximately given by 
$v_\text{F}\sim10^6$ m/s. 
In this notation, $\boldsymbol{\sigma}$ and $\boldsymbol{\tau}$ are vectors composed of $2$$\times$$ 2$ Pauli matrices and refer to spin space and sublattice space, respectively. 
The honeycomb lattice of graphene is composed of
`A' and `B' sublattices, as 
illustrated  in Fig.~\ref{fig1}. The strength of 
Rashba SOC, magnetization, and chemical potential are labeled by $\alpha$, $\textbf{h}$, and $\mu$, respectively. In our calculations that follow, we have considered an arbitrary direction for the magnetization orientation so that $\textbf{h}=(h_x,h_y,h_z)$. The magnetization might be induced into the graphene layer by a magnetized substrate such as YIG or the 
application of an external magnetic field \cite{exp_f1,ex2,ex3}. The intrinsic staggered SOC is negligibly small in a typical graphene layer, where the ratio of extrinsic to intrinsic SOC is on the order of 100 \cite{H.Min,J.C.Boettger}. However, since 
the magnetization $h_z$ can play the same role as this type of SOC \cite{RMP-2009-Neto,S.Sarma}, 
 its presence can be revealed by subtle features  in the  response of graphene
 to RH and LH polarized waves. Thus, the overall effect on the results is  similar to 
 having  $h_z$  present, even in situations where the intrinsic SOC is considerably large. 
 The key features due the presence of magnetization are described below in detail. Hence, 
the associated field operators have four components, carrying
the spin and sublattice degrees of freedom, i.e., 
$\Psi^\dagger(\textbf{r})=(\psi_{\text{A}\uparrow}^\dagger,\psi_{\text{B}\uparrow}^\dagger,\psi_{\text{A}\downarrow}^\dagger,\psi_{\text{B}\downarrow}^\dagger)$ and $\Psi(\textbf{r})=(\psi_{\text{A}\uparrow},\psi_{\text{B}\uparrow},\psi_{\text{A}\downarrow},\psi_{\text{B}\downarrow})^T$.

The components of the frequency-dependent permittivity 
tensor $\epsilon_{ab}(\omega)$ 
are related to the components of 
the conductivity tensor  $\sigma_{ab}(\omega)$ through the standard relations,
\begin{subequations}
\begin{align}
&\epsilon_{ab}(\omega) = \delta_{ab} +i \frac{\sigma_{ab}(\omega)}{\epsilon_0\omega}, \label{eps_ab}\\
&\sigma_{ab}(\omega) = \frac{i}{\omega} \lim_{\textbf{q}\rightarrow 0}\Big\{\Pi_{ab}(\omega,\textbf{q}) - \Pi_{ab}(0,\textbf{q})\Big\}.\label{optcon}
\end{align}
\end{subequations}
Here $\delta_{ab}$ is the Kronecker delta,
the indices $a,b$ run over $x,y$,  and  $\epsilon_0$ is the permittivity of free space. 
The current-current correlation function is given by
\begin{align}\label{sig_ab}
&\Pi_{ab}(\omega,\textbf{q}) = \frac{e^2}{\hbar}\int \frac{d\textbf{k}}{(2\pi)^2}\sum_{n}\sum_{\tau=A,B}\sum_{s=\uparrow\downarrow} \times {\cal F}(\varepsilon_n,\omega,\mu)\nonumber\\&\text{Tr}\Bigg\{  {\cal J}_{a,\tau,s} G_{\tau,s}\Big(\varepsilon_n+i\omega,\textbf{k}+\textbf{q}\Big){\cal J}_{b,\tau,s} G_{\tau,s}\Big(\varepsilon_n,\textbf{k}\Big)\Bigg\}.
\end{align} 
Here $\tau,s$ are sublattice and spin degrees of freedom, respectively, and the components of the current operator are ${\cal J}_{a,\rho,s}$. The Fermi-Dirac function $f(X)$ determines the temperature dependency ($T$) of optical conductivity and permittivity
\begin{subequations}
\begin{align}
&{\cal F}(\varepsilon_n,\omega,\mu) = f(\varepsilon_n-\mu) - f(\varepsilon_n+\omega-\mu),\\
& f(x) = \frac{1}{e^{\beta x}+1},\;\; \beta=\frac{1}{k_BT},
\end{align} 
\end{subequations}
in which $k_B$ is the Boltzmann constant.

\section{results and discussions}\label{sec:results}

We first present results for
the optical conductivity and 
analyze the physical origins of its various features in Sec.~\ref{sec:optcond}. 
In Sec.~\ref{sec:hand}, the
components of the dispersive permittivity tensor are studied as a
function of chemical potential. Next,  
the effects of frequency, chemical potential, 
and magnetization on the circular dichroism 
 of the device shown in Fig.~\ref{fig1} will be presented.

\subsection{Finite-frequency optical conductivity and physical analysis}\label{sec:optcond}

To obtain the permittivity tensor with frequency dispersion,
the components of the Green's function used in Eq.~(\ref{sig_ab}) need to be derived. 
For concreteness, and 
to simplify the expressions, we set $\textbf{h}=(0,0,h_z)$ in what follows. Nonetheless, we have obtained
 expressions for generic cases with  $\textbf{h}\neq 0$. Using the low-energy Hamiltonian (\ref{hamil}), the components of the Green's function in the presence of
 an  exchange field $h_z$, and SOC are expressed by,
\begin{subequations}
\begin{flalign}
\Omega G_{11}= &- 4 (h_z + i\omega) \alpha^2 \nonumber &&\\
&-(h_z - i\omega) (h_z - \textbf{k} + i\omega) (h_z + \textbf{k} + i\omega),
\end{flalign} 
\begin{flalign}
\Omega G_{12}= +2i(k_x-ik_y)(h_z + i\omega)\alpha, &&
\end{flalign} 
\begin{flalign}
\Omega G_{13}= +(k_x-ik_y)(h_z - \textbf{k} + i\omega)(h_z + \textbf{k} + i\omega), &&
\end{flalign} 
\begin{flalign}
\Omega G_{14}= +2i(k_x-ik_y)^2\alpha, &&
\end{flalign} 
\begin{flalign}
\Omega G_{22}= +(h_z - \textbf{k} - i\omega) (h_z + \textbf{k} - i\omega) (h_z + i\omega), &&
\end{flalign} 
\begin{flalign}
\Omega G_{23}= -2i(h_z  - i\omega) (h_z  + i\omega)\alpha, &&
\end{flalign} 
\begin{flalign}
\Omega G_{24}= +(k_x-ik_y)(h_z + \textbf{k} - i\omega)(h_z - \textbf{k} - i\omega), &&
\end{flalign} 
\begin{flalign}
\Omega G_{33}= +(h_z - \textbf{k} + i\omega) (h_z + \textbf{k} + i\omega) (h_z - i\omega), &&
\end{flalign} 
\begin{flalign}
\Omega G_{34}= 2i(k_x-ik_y)(h_z - i\omega)\alpha, &&
\end{flalign} 
\begin{flalign}
\Omega G_{44}= &+ 4 (h_z - i\omega) \alpha^2 \nonumber &&\\
&+(h_z + i\omega) (h_z - \textbf{k} - i\omega) (h_z + \textbf{k} - i\omega),
\end{flalign} 
\begin{flalign}
\Omega =&\Big( i\omega +\big[ {\cal C} -{\cal D}\big]^\frac{1}{2} \Big)\Big( i\omega -\big[ {\cal C} -{\cal D}\big]^\frac{1}{2} \Big)\times\nonumber &&\\
&\Big( i\omega +\big[ {\cal C} +{\cal D}\big]^\frac{1}{2} \Big)\Big( i\omega -\big[ {\cal C} +{\cal D}\big]^\frac{1}{2} \Big), &&
\end{flalign} 
\begin{flalign}
&{\cal C} = h_z^2 + \textbf{k}^2 + 2\alpha^2,&&\\
&{\cal D} =2\sqrt{h_z^2\textbf{k}^2+\textbf{k}^2\alpha^2 +\alpha^4}.&&
\end{flalign} 
\end{subequations}
The other components of the
Green's function can be inferred
 from  symmetry arguments. 
 Substituting the Green's function components into Eq.~(\ref{optcon}), 
 we find the following expression for the dynamical $\sigma_{xx}(\omega)$ and $\sigma_{xy}(\omega)$ components of optical conductivity:
\begin{align}\label{sigxx_yx}
&\sigma_{xx(xy)}(\omega,\mu)=\frac{2\pi^2e^2}{\hbar}\int \frac{d\textbf{k}}{(2\pi)^2}\int \frac{d\varepsilon_n}{2\pi} {\cal F}(\varepsilon_n,\omega,\mu)\times \nonumber \\
\Big\{& \pm{\cal A}_{11}(\varepsilon_n^+){\cal A}_{33}(\varepsilon_n) \pm {\cal A}_{22}(\varepsilon_n^+){\cal A}_{44}(\varepsilon_n)\nonumber \\
&+ {\cal A}_{33}(\varepsilon_n^+){\cal A}_{11}(\varepsilon_n) + {\cal A}_{44}(\varepsilon_n^+){\cal A}_{22}(\varepsilon_n)\nonumber \\
&+ {\cal A}_{34}(\varepsilon_n^+){\cal A}_{21}(\varepsilon_n) + {\cal A}_{43}(\varepsilon_n^+){\cal A}_{12}(\varepsilon_n)\nonumber \\
&\pm {\cal A}_{12}(\varepsilon_n^+){\cal A}_{43}(\varepsilon_n) \pm {\cal A}_{21}(\varepsilon_n^+){\cal A}_{34}(\varepsilon_n)
\Big\},
\end{align} 
where $\varepsilon_n^+ = \varepsilon_n + \omega$, and the 
 definitions of ${\cal A}_{ij}$ are given in Appendix \ref{appx:GF}. Here
 the symbol $\pm$ refers to the
 $xx(xy)$ indices ($+(-)$), respectively.
 It is evident that
 owing to the complexities of these expressions, solutions can only be obtained  numerically.

\begin{figure}[t!]
\includegraphics[width=0.4\textwidth]{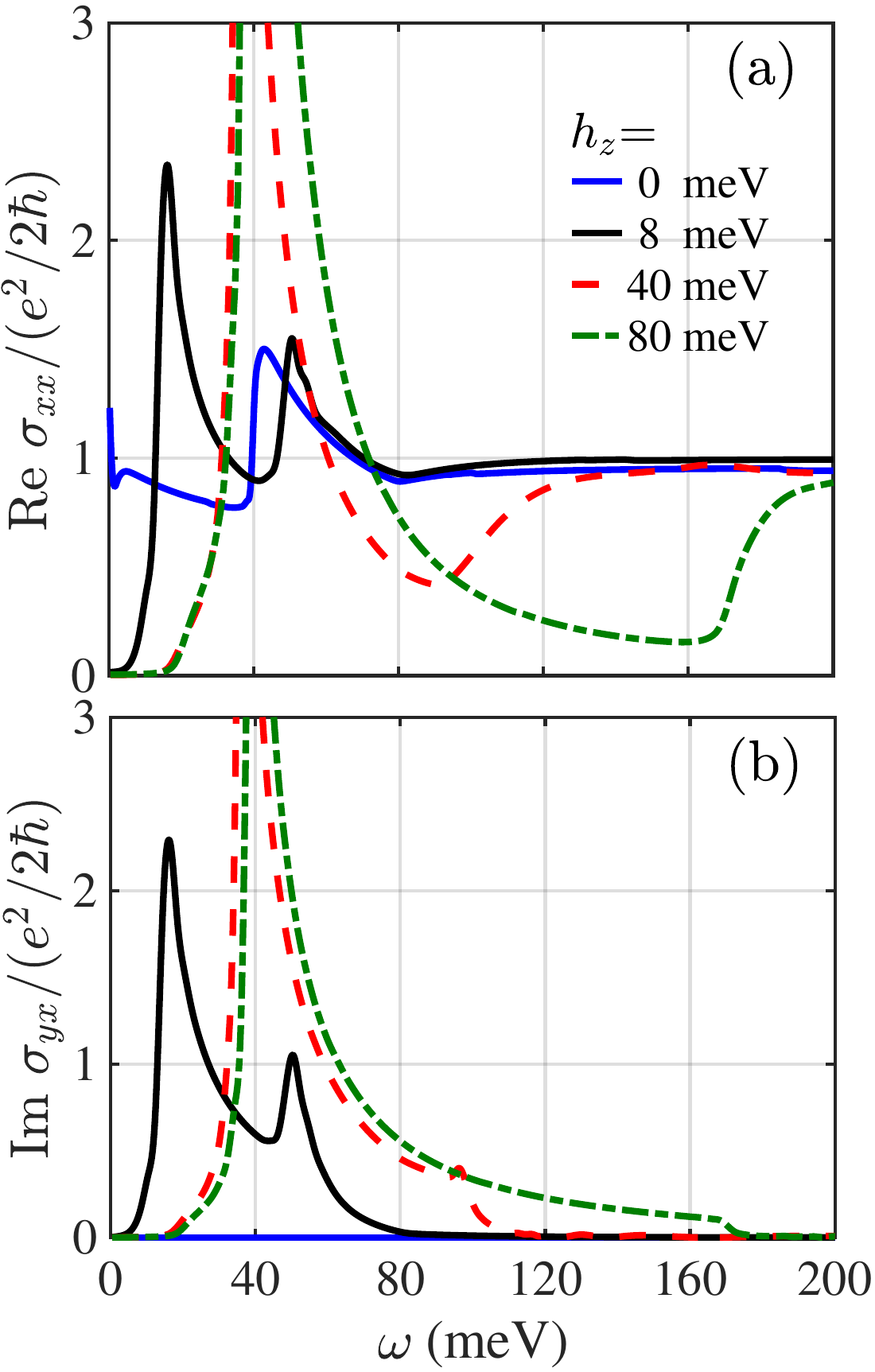}
\caption{\label{fig2} (Color online). The components of the
optical conductivity as a function of frequency. (a) and (b) are the real part of $\sigma_{xx}(\omega)$ and imaginary part of $\sigma_{yx}(\omega)$, respectively. The magnetization is oriented along the $z$ axis, i.e., $\textbf{h}=(0,0,h_z)$. The chemical potential is set to zero, $\mu=0$, and various values of magnetization 
strength are considered: $h_z=0, 8, 40, 80$ meV. The strength of the SOC is set  to $\alpha=20$ meV.}
\end{figure}

\begin{figure*}[t!]
\includegraphics[width=\textwidth]{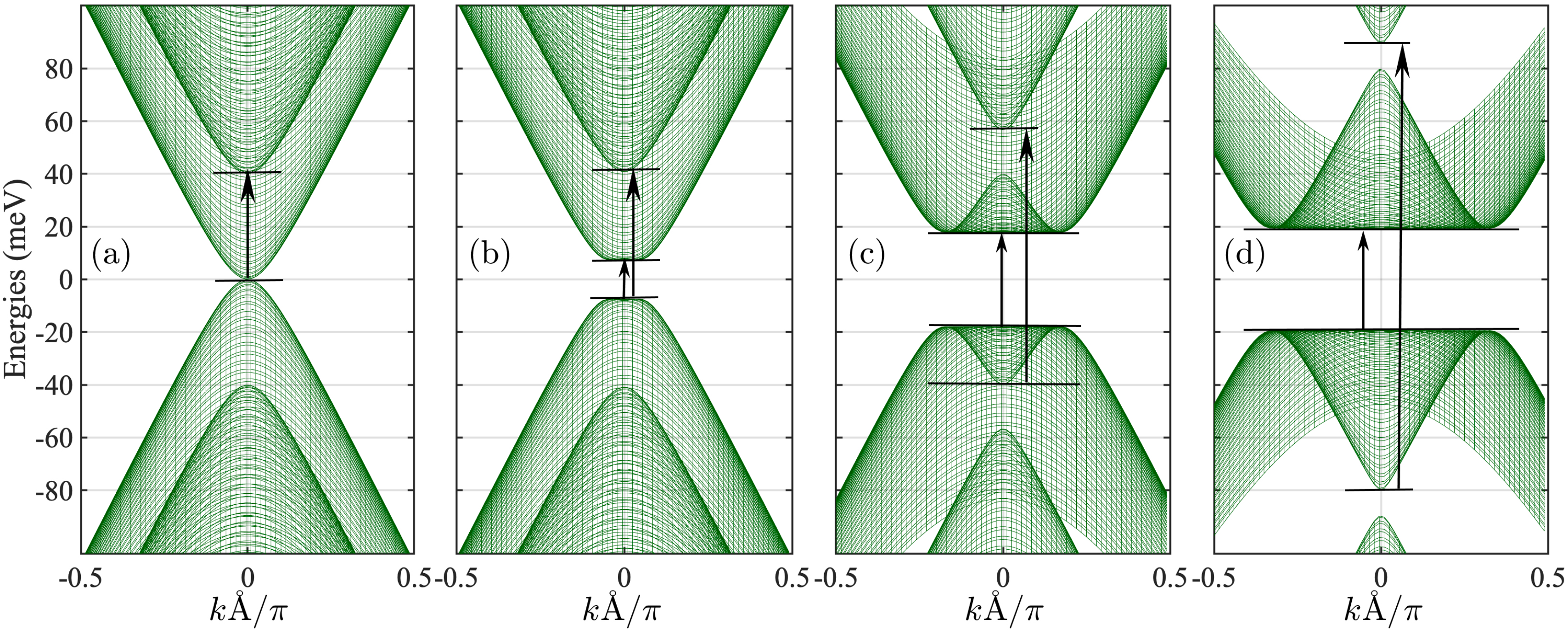}
\caption{\label{fig3} (Color online). The band structure of graphene in the presence of SOC and magnetization
plotted as a function of the normalized 
momentum $k {\textup \AA}/\pi$. The chemical potential is set to zero;
 $\mu=0$, and the strength of the SOC is fixed at $\alpha=20$~meV. 
 The magnetization is oriented along the $z$-axis and its magnitude increases 
 (from
  left to right) with the following values (in meV): (a) $h_z=0$, (b) $h_z=8$, 
  (c) $h_z=40$, and (d) $h_z= 80$.}
\end{figure*}

In Fig.~\ref{fig2}, the real part of the 
longitudinal optical conductivity $\sigma_{xx}(\omega)$ and imaginary 
part of the transverse optical conductivity $\sigma_{yx}(\omega)$ 
are plotted as a function of the incident photon frequency $\omega$. To this end, we have evaluated Eqs.~(\ref{sigxx_yx}), considering a situation where the magnetization is oriented along the $z$ axis, perpendicular to the plane of the graphene sheet, shown in Fig.~\ref{fig1}. 
For this particular system,  the following relations hold: 
$\sigma_{yy}(\omega)=\sigma_{xx}(\omega)$ and $\sigma_{xy}(\omega)=-\sigma_{yx}(\omega)$.
The imaginary and real parts of $\sigma_{xx}(\omega)$ and $\sigma_{yx}(\omega)$ can be obtained by the Kramers-Kronig relationship. 
The components of the 
conductivity are normalized by the conductance unit $e^2/2\hbar$. 
To facilitate the analysis of the optical conductivity, the chemical potential is set to zero, $\mu=0$. 
Nevertheless, later the chemical potential shall be nonzero for illustrating its influence
on the relevant material parameters. 
The strength of the 
SOC is set to $\alpha=20$~meV, 
consistent with inferred 
values 
from
experiments \cite{J.Balakrishnan}, 
although our conclusions depend on
 the chemical potential and magnetization strength relative to $\alpha$. 
To perform numerically stable calculations, 
we have used the Lorentz model for the Dirac delta functions with  
narrow width,  $\eta=0.01$~meV.
We have also set
the  temperature to $T=0.01$~K 
in all subsequent calculations.

When $h_z=0$~meV in Fig.~\ref{fig2}, the transverse conductivity $\sigma_{yx}(\omega)$ vanishes. The longitudinal conductivity $\sigma_{xx}(\omega)$ shows a weak Drude response at very low frequencies, $\omega\rightarrow 0$. 
It is evident that  $\sigma_{xx}(\omega)$ 
peaks at $\omega\approx 40$~meV and reaches the universal 
background conductivity, $\sigma_0\equiv e^2/2\hbar$, at
 higher frequencies. 
 When the 
magnetization is increased to a nonzero value, $h_z=8$~meV, 
 two peaks in $\sigma_{yx}(\omega)$ emerge at 
 $\omega\approx 20$~meV and $\omega\approx 50$~meV,  then decay at higher frequencies.
 The  peaks 
 observed  in $\sigma_{yx}(\omega)$
 appear at the same frequencies 
 for  the longitudinal $\sigma_{xx}(\omega)$,
and the
 conductivity 
 approaches  
its background value  for $\omega \gtrsim 80$~meV. 
 Also, as seen, both the longitudinal and transverse 
 conductivities are zero for small frequencies, $0\lesssim\omega\lesssim8$~meV. 
 Increasing 
 the magnetization to larger values, e.g., $h_z=40$~meV
 and $h_z=80$~meV, 
 both the transverse and longitudinal components show that 
 the magnitude of the first peak increases 
 considerably whereas the second peak dampens out. 
 The $\sigma_{xx}(\omega)$ component shows a steep decline
 %decreasing behavior between 
 from 
 $\omega\approx 40$~meV 
 for both $h_z=40$~meV
 and $h_z=80$~meV,
 before increasing again
 at 
$\omega\approx 90$~meV 
and  $\omega\approx 170$~meV , respectively.
Increasing the 
frequency higher results again in 
the longitudinal conductance leveling off at
 $e^2/2\hbar$. 
Another important effect of the magnetization 
that can play a role in 
practical device applications
is that 
within the low-frequency regime, 
 the width of the zero conductance
 region  can  increase 
 by increasing the magnetization strength
 to a threshold value.

In order to gain further  insight into the physical origins of the optical conductivity 
results 
above, we plot the associated band structures as a function of momentum $k$ in Fig.~\ref{fig3}. 
For consistency, the  values for the chemical potential and  SOC
strength  are 
the same as those in Fig.~\ref{fig2}. 
In Figs.~\ref{fig3}(a)-\ref{fig3}(d), the magnetization increases as
$h_z=0, 8, 40, 80$~meV, respectively. 
As seen in Fig.~\ref{fig3}(a), in the presence of SOC, the bands associated with different spins
 are split by the amount $2\alpha=40$~meV at $k=0$. Also, the valence 
 band and conduction band just touch at $k=0$ and $E=0$. 
 The latter results in weak intraband transitions and a Drude response at $\omega=0$~meV, 
 as observed in Fig.~\ref{fig2}(a). The interband transitions occur at energies 
  $\omega\gtrsim 40$~meV 
  (shown by the
   arrow in Fig.~\ref{fig3}(a)) 
   and show up in 
   the conductivity (Fig.~\ref{fig2}(a)) as a peak at around $\omega\approx 40$~meV. 
   At high enough energies, the transition rate slows
    until finally reaching a constant rate
     equivalent to the universal conductivity $e^2/2\hbar$.
   When   the magnetization 
   is increased to  $h_z=8$~meV [Fig.~\ref{fig3}(b)], a small energy gap ($\approx 8$~meV) 
   opens up in the band structure at 
   the Fermi level, $E=0$. 
   There is also a slight shift upward
   of the second valence band. 
   The interband transitions shown by the small and large arrows
   are the origins of the
     two peaks in 
     the optical conductivity at approximately 
     $16$~meV and $50$~meV, as seen in Fig.~\ref{fig2}. 
     The small gap ($\approx 16$ meV) between the bottom of 
     the conduction band and
     the  top of the valence band prevents any transitions,
      and hence results in zero conductivity for frequencies less than 
   $\omega\lesssim16$~meV [see Fig.~\ref{fig2}]. 
   
   Upon increasing the magnetization further to $h_z=40$~meV [Fig.~\ref{fig3}(c)], the band gap 
     widens to approximately  $35$~meV, 
   and the conduction band around $k=0$ acquires a
   dome-like segment with
    its top at  $40$~meV from $E=0$~meV. 
    The same feature occurs,
    but inverted, in the valence band. 
    As shown by the small and large arrows, 
    two types of interband transitions can 
    be expected at $\omega \approx 35$ meV and $\omega \approx 100$ meV, respectively. 
   These band structure transitions correlate with 
   discernible features in the optical conductivity
    shown in  Fig.~\ref{fig2}. 
    Finally, increasing the  magnetization to $h_z=80$~meV [Fig.~\ref{fig3}(d)] results in a band structure with similar features to those of $h_z=40$~meV shown in Fig.~\ref{fig3}(c). Therefore, the associated optical conductivities are also similar. Another main feature found in both components of 
    the optical conductivity
    is that when increasing $h_z$,  there is a
     considerable enhancement of the first peak. 
     This can be understood by
     considering the corresponding band structures in Figs.~\ref{fig3}(c) and \ref{fig3}(d),
     and comparing  to Figs.~\ref{fig3}(a) and \ref{fig3}(b), respectively.
      The 
      bottoms  of the  valence and conduction bands  become flattened and
      extended  in Figs.~\ref{fig3}(c) and \ref{fig3}(d),
      thus
     providing many more available states for interband transitions
    (as  indicated by the small arrows).

\subsection{Handedness switching and dielectric response }\label{sec:hand}

\begin{figure}[!t] 
\centering
\includegraphics[width=0.48\textwidth]{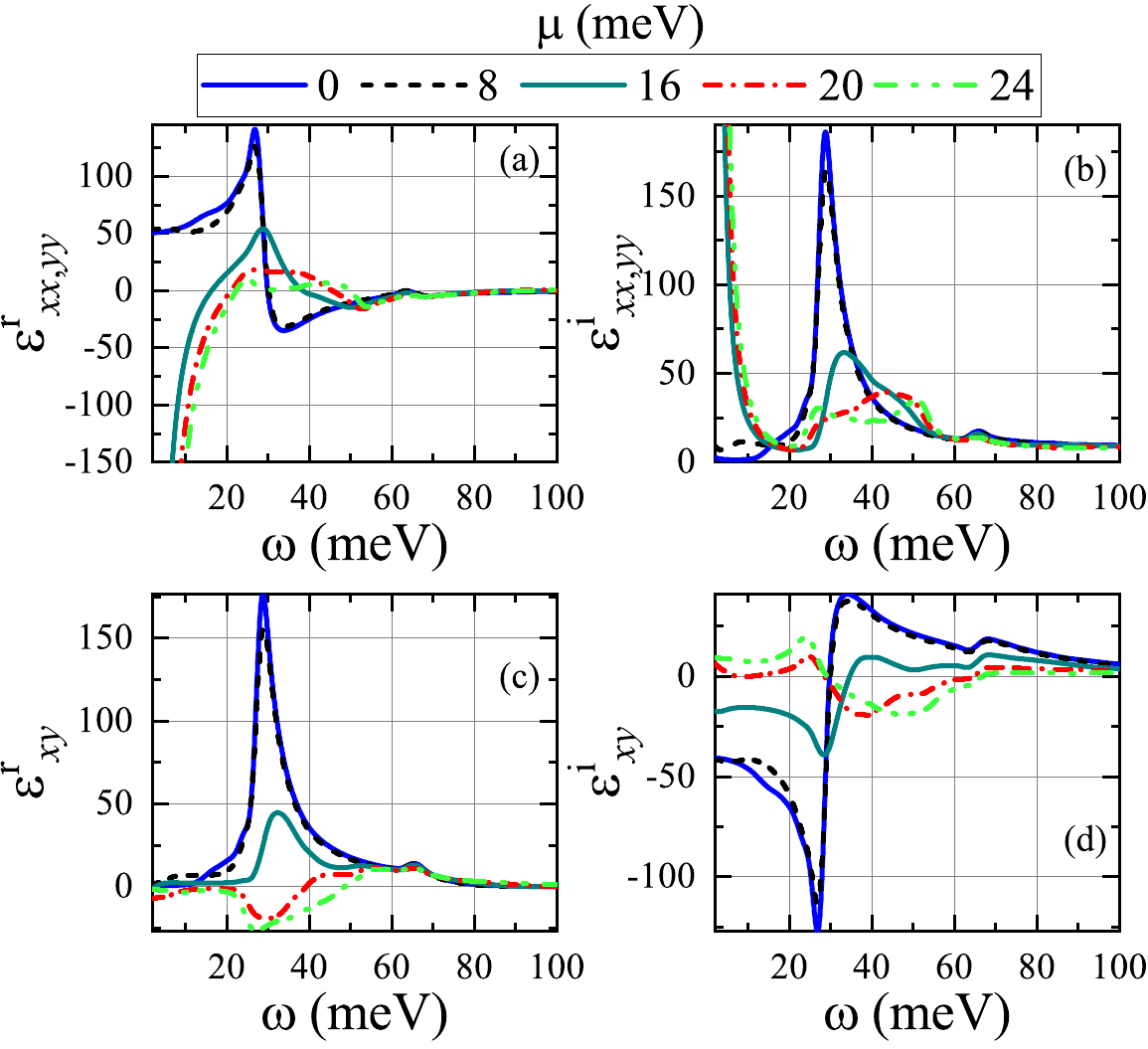}
\caption{Permittivity tensor components for the graphene sheet with finite SOC and magnetization. Both the real and imaginary components
are shown for several different chemical potentials $\mu$. The 
Zeeman field is set to $h_z=20$~meV.
From symmetry considerations, $\epsilon_{xy}(\omega)=-\epsilon_{yx}(\omega)$, 
the other off-diagonal permittivity components can be deduced.
}
\label{fig4}
\end{figure}

As seen in Fig.~\ref{fig2}, the interplay of 
Rashba SOC and a
magnetization perpendicular to the graphene film
results in a 
strongly modified
longitudinal optical conductivity,
and generation of
a  finite transverse conductivity.
The 
 corresponding  components of the 
 permittivity tensor for graphene, 
 $\rttensor{\epsilon}_1 (\omega)$,
can  thus be expressed  as:
\begin{eqnarray}
\rttensor{\epsilon}_1(\omega)
=\left( \begin{array}{ccc}
\epsilon_{xx}(\omega) & \epsilon_{xy}(\omega) & 0 \\
\epsilon_{yx}(\omega) & \epsilon_{yy}(\omega)  & 0 \\
0 &0 &1  \end{array}\right).
\end{eqnarray}
In Fig.~\ref{fig3},
the
behavior of the diagonal and off-diagonal 
components of $\rttensor{\epsilon}_1 (\omega)$ are shown
as a function of frequency.
A representative set of parameter values are considered with
 the magnetization strength set at $20$~meV,
 and  $\mu$
 varies from 
 %0, 8, 16, 20, 24$~meV.
 $0$-$24$~meV.
 The chemical potential is controllable via a gate voltage, as shown in 
 Fig.~\ref{fig1}.
 As discussed earlier and illustrated in, e.g., Fig.~\ref{fig4}(c), 
 for cases with $\mu=0$ and $\mu=8$~meV, 
 the Fermi energy resides inside the gap of 
 the  
 band structure, and therefore
 only the interband transitions are allowed. 
 For larger values of the
chemical potential, i.e., $\mu=16, 20, 24$~meV, the 
intraband transitions are additionally allowed. 
The interband transitions 
are responsible for the Drude-like response at low frequencies in 
the longitudinal components of 
the
dielectric response, i.e., $\epsilon_{xx,yy}(\omega)$. 
The Drude-like  response
to an electromagnetic wave
 can be clearly seen in Fig.~\ref{fig4}(a) and 
Fig.~\ref{fig4}(b)  
as
 $\omega\to 0$. 
 Considering the 
 previous analysis of the components of 
 the
 optical conductivity,
 which showed a strong frequency dependence 
 for finite values of $h_z$, it is evident that the large variations
  in the permittivity components are also
  strongly influenced by the presence of magnetization. 
  Moreover, the
 frequency  of the  first peaks in 
  Fig.~\ref{fig2} is directly related to the 
   magnitude of the 
   magnetization induced into graphene. As shown below,
   the strong transverse dielectric response
   seen in 
   Figs.~\ref{fig4}(c)-\ref{fig4}(d), which reveals the presence of an
   extrinsically induced SOC in the graphene sheet (with no extrinsic SOC, the components 
   $\varepsilon_{xy(yx)}$ vanish),
   is crucial for 
   circular dichroism and polarization control.  

Employing the computed frequency-dependent
 permittivity and conductivity tensors for various parameter sets, 
 one now can easily study the absorption
of EM waves from a hybrid 
device containing
 magnetized spin-orbit coupled graphene, such as
the layered configuration shown in 
 Fig.~\ref{fig1}. This simple device 
 consists of graphene sheet
 adjacent to a 
 dielectric spacer layer with 
 SOC (and possibly magnetization).
 The substrate consists
 of a reflective ground plate, which 
 can be served by a metal, which will eventually 
 be assumed to have  perfect conductivity. 
 The electric field component of the
 normally  incident EM
 wave in the vacuum region 
 is polarized in the $xy$ plane,
 and we consider a circular polarization, so that the incident EM field 
 components  ${\bm E}_{0i}$
 are out of phase, i.e.,
  ${\bm E}_{0i}=(1,\pm i)$, for right-handed ($-$),
or left-handed ($+$) circular polarization.
When determining how much of the incident EM energy is absorbed by the structure in 
Fig.~\ref{fig1},
we invoke  Maxwell's equations. The equations, specific fields, and boundary conditions used for the results presented in the following can be found in Appendix \ref{appx:MAXWELL}. 

The fraction of energy that is absorbed by the 
system is determined by the absorptance $A(\omega)$:
$A(\omega)=1-T(\omega)-R(\omega)$,  
where $T(\omega)$ is the transmittance and $R(\omega)$ is the reflectance,
consistent with energy conservation.
In determining the absorptance  of the graphene system, we
consider the time-averaged Poynting vector in the direction 
perpendicular to the interfaces (the $z$ direction), 
${\bm S}(\omega) {=} \Re{\{{\bm E}(\omega)\times{\bm H}^*(\omega) \}}/2$.
We now take the limit of
 a metallic substrate, so 
 there is no transmission of EM fields,  and $T=0$,
 and the tangential electric field at the spacer/metal boundary vanishes.
Upon inserting the electric and magnetic fields
for the vacuum region, we find,
\begin{align}
A (\omega)= 1 - \frac{|{\bm E}_{0r}(\omega)|^2}{|{\bm E}_{0i}(\omega)|^2}.
\end{align}
Here we have normalized energy relative to the incident plane wave energy $S_0$,
where 
$S_0 = (1/\eta_0) |{\bm E}_{0i}|^2$.
The reflection coefficients ${\bm E}_{0r}$ are  found upon 
using conditions (\ref{Econd})-(\ref{Hcond}). To quantify the effect that  graphene has on the handedness of circularly polarized light, the quantity $\Psi(\omega)$ is introduced:
\begin{align}
\Psi (\omega)= \frac{A_{-}(\omega) - A_{+}(\omega)}{A_{+}(\omega)+ A_{-}(\omega)},
\end{align}
which describes the amount of circular dichroism
 through the difference in  absorption of LH ($+$) and RH ($-$) circularly polarized 
EM waves.
Therefore, $\Psi (\omega)>0$ indicates dominant right-handed
 absorption, while
 $\Psi (\omega)<0$, indicates  that
 left-handed polarization tends to be absorbed more.

\begin{figure}[!b] 
\centering
\includegraphics[width=0.49\textwidth]{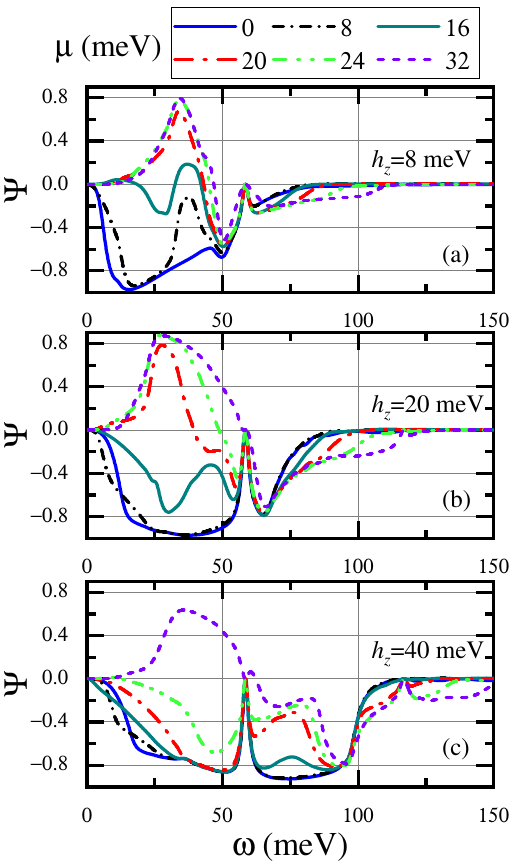}
\caption{The circular dichroism $\Psi(\omega)$ as a function of frequency. Several out-of-plane Zeeman fields $h_z$ are considered (as shown).
Through variations in the gate voltage, the reflected EM energy can exhibit dominant right-handed (positive curves)
or left-handed (negative curves) behavior.
}
\label{fig5}
\end{figure} 

To evaluate the circular dichroism and handedness characteristics of the device shown in 
Fig.~\ref{fig1}, the thickness of the spin-orbit coupled insulator layer (the yellow layer in Fig.~\ref{fig1}) 
is set to a representative value of $5$~$\mu$m, and the medium is assumed nondispersive.
The latter assumption can be easily achieved with a  large band gap semiconductor alloy, 
involving heavy elements to support SOC as well. The 
CD factor, $\Psi(\omega)$,
is 
shown in Fig.~\ref{fig5}
as a function of
 the frequency of
 the  incident EM wave. 
The legend shows the 
chemical potentials that are considered,
with $\mu$ ranging from 0 to 32~meV, 
ensuring 
the most pertinent cases are shown.
Additionally, each panel in Figs. \ref{fig5}(a)-\ref{fig5}(c) considers a different finite value of the longitudinal magnetization $h_z$, with
 $h_z=8,20,40$~meV, respectively.
 Beginning with Fig.~\ref{fig5}(a),
 the results illustrate that
at
 the charge neutrality point, $\mu=0$, and moderate magnetization strength,
  there is strong
  circular dichroism,  with the relative absorption favoring  left-handed EM
  waves  for a broad range of frequencies ($\Psi(\omega)<0$).
  Increasing the chemical potential to $\mu=8$~meV yields larger variations 
  in $\Psi(\omega)$ and a narrower frequency window for strong left-handed
  absorption. When the chemical potential is set to $\mu=16$ meV,
  $\Psi(\omega)$ gets shifted upward and oscillates about zero
    for frequencies lower than $\omega \lesssim 50$~meV, resulting in
left and right handedness switching as a function of frequency.
Further increasing $\mu\gtrsim 20$ meV,
results in  $\Psi(\omega)>0$ 
 for
$\omega \lesssim 45$ meV. 
At higher frequencies,
$\Psi(\omega)<0$, and levels off toward unity as
$\epsilon_{xy}(\omega)$ vanishes [see Fig.~\ref{fig4}],
with  both left-handed and right-handed polarizations 
 absorbed equally.
In Figs.~\ref{fig5}(b) and \ref{fig5}(c),
increasing the magnetization is shown
to have a profound effect on the gate-controlled handedness-switching.
This is seen in  Fig.~\ref{fig5}(b), where 
$h_z=20$ meV, and the optical response demonstrates 
an effectively larger  circular dichroism over a wider frequency range.
The results show that the
relative absorption of left-handed and right-handed circularly polarized
waves with these parameters
can be manipulated  through 
variations  in frequency and chemical potential.
Doubling the magnetization to $h_z=40$ meV [Fig.~\ref{fig5}(c)] has a severe impact on the effectiveness of the device
for handedness-switching  of  EM waves.
Although
at this larger magnetization the structure now exhibits circular dichrosim over
a broader range of frequencies, 
$\Psi(\omega)$ is always negative or zero, except for
the largest chemical potential, $\mu=32$ meV, 
where $\Psi(\omega)$ is positive (RH dominant)
for $0\leq \omega \lesssim 60$~meV.
For all other values of $\mu$ shown,
the 
LH  polarization
state  
always dominates  ($\Psi(\omega)<0$).

\begin{figure}[!t] 
\centering
\includegraphics[width=0.49\textwidth]{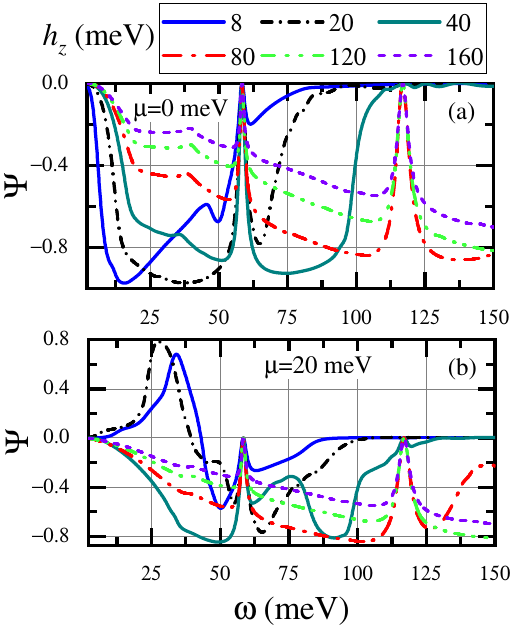}
\caption{(Color online).  The circular dichroism factor $\Psi(\omega)$ against the frequency of 
the incident  
circularly polarized EM wave. 
A broad range of 
magnetizations $h_z$ at two different values of chemical potential are considered: $\mu=0,20$~meV.
}
\label{fig6}
\end{figure}

To explicitly show the influence of 
magnetization on CD, 
the $\Psi(\omega)$ factor against frequency for 
various values of $h_z$ is plotted in Fig.~\ref{fig6}. 
Two 
values of the
chemical potential are considered: 
(a) $\mu=0$ and (b) $\mu=20$~meV. 
As is clearly seen in Fig.~\ref{fig6}(a), 
increasing the magnetization does not induce a sign change in
$\Psi(\omega)$, and hence there is no handedness-switching
at the 
charge neutrality point, $\mu=0$.
Nonetheless, 
there are strong variations in $\Psi(\omega)$ and a dominant  LH
response over a frequency window that widens with increasing $h_z$.
Once the chemical potential is shifted away
 from the neutrality point, e.g., $\mu=20$~meV, 
 Fig.~\ref{fig6}(b) illustrates that  now 
 $h_z$ can induce
  handedness-switching. 
  This switching is accessible 
  for frequencies where
  $\omega\lesssim~50$~meV and 
  magnetizations
  $h_z=8,16$~meV. 
This  trend  has been found to continue for larger
    chemical potentials,
  whereby 
    manipulating  $h_z$ can increase the impact in
  calibrating and controlling the handedness-switching (not shown).
Another feature that can be  seen in Figs.~\ref{fig5} and \ref{fig6} 
is the vanishing 
$\Psi(\omega)$ factor at specific frequencies. 
This follows from Eq.~(\ref{reflect1})
and Eq.~(\ref{reflect2}), 
where the reflectivity coefficients, $|E^\pm_{0rx}|^2=|E^\pm_{0ry}|^2=1$ when
$k_2 d = n \pi$ (for $n=1,2,\ldots$). This is equivalent to having the
spacer layer at the resonance width  $n \lambda/2$,
where  destructive interference
occurs  from reflected waves at both edges of the insulator layer ($\lambda$ is the wavelength of light in the insulator).
This perfect reflectivity  is most pronounced in 
Figs.~\ref{fig5}(b)-\ref{fig5}(c), and Fig.~\ref{fig6}. 
From the resonance condition above,
 $\Psi(\omega)$ vanishes at $\omega~({\rm meV})\approx 1240 n/(2 d \sqrt{\epsilon_2}) \approx 60, 120,\ldots$
 ($d$ is given in microns).
Therefore, these specific frequencies depend solely on the material and geometrical characteristics of the
spacer layer, and are independent of the dielectric response of the magnetized spin-orbit coupled graphene. Controlling the handedness of circularly polarized light with an external field offers several interesting possibilities for device applications. Through extensive parameter sweeps, general relationships between the magnetization, chemical potential, and frequency can be found at the crossover point, which can be beneficial for device fabrication. Although such an effort is outside the scope of this paper, we defer this interesting topic as a project for future works.       

Note that 
there is no circular dichroism 
 when the transverse components of the dielectric response are absent,
 i.e., $\epsilon_{xy,yx}(\omega)=0$. Therefore, a nonzero CD factor reveals
 important signatures involving 
  the interplay of SOC 
 and magnetization. Since the chemical potential of graphene is easily tunable by a gate voltage, the 
 explored handedness-switching also 
 demonstrates the extrinsically induced SOC and magnetization in graphene. 
 On the technological side, this
  handedness-switching can be exploited for devising optoelectronic applications
 including  chemical and biological sensors,
 where  the corresponding polarization- and frequency-dependent  absorption signatures 
 can be controlled externally, e.g., by application of a gate voltage or 
 magnetic field. The well-defined peaks in the
 absorption signatures were shown to provide a way to characterize the extrinsic SOC
 without ambiguity, as evidenced in the handedness-switching phenomenon and anomalous conductivity. Indeed, the magnitude of the extrinsic SOC can be estimated by the peaks in Fig.~\ref{fig2},
 as discussed above.
 If there is no magnetization, $h_z=0$,
the externally induced anomalous conductivity and handedness-switching disappears.
Since the intrinsic SOC plays the same role as $h_z$, its magnitude can be revealed 
using the same characterization techniques used above when the magnetization was present.               

\section{conclusions}\label{sec:concl}

In summary, we have studied the 
frequency-dispersive 
 optical conductivity and associated dielectric response of 
 magnetized graphene with Rashba spin-orbit coupling (SOC). 
 Our results revealed that the strength and type of SOC can be unambiguously concluded and 
 estimated through measurements of the frequency dependence of the optical conductivity. The band structure analysis
  illustrated that SOC and magnetization can
 generate  two well-separated peaks in the conductivity, thus
 determining the strength of the SOC and magnetization. The transverse Hall conductivity, due to the 
 interplay of magnetization and SOC, results in a nonzero circular dichroism. Exploring the permittivity  tensor 
 and conductivity, we studied the absorptance of a  simple device, consisting of  magnetized spin-orbit coupled graphene on top of an insulator layer 
 and  perfectly conducting metal substrate. Our findings showed
  that both the
  magnetization and chemical potential offer practical
  control mechanisms  for  handedness-switching and circular dichroism effects,
   where there is tunable absorption of left-handed and right-handed circularly polarized light. In addition, these findings offer a possible physical mechanism for recent experiments \cite{C.Hwang}. Accordingly, the interaction of graphene and a $\rm SiC$ substrate 
   can be the source of  SOC and magnetism in these platforms.      

\acknowledgements
Part of the calculations were performed using HPC resources from the DOD High Performance Computing Modernization Program (HPCMP). K.H. is supported in part by the NAWCWD In Laboratory Independent Research (ILIR) program and a grant of HPC resources from the DOD HPCMP.

\appendix

\section{Green's functions and definition of parameters}\label{appx:GF}
In this appendix, we present the variables defined in the main text. Specifically, the variables ${\cal A}_{ij}$, $i,j\equiv 1,2,3,4$, in Eq. (\ref{sigxx_yx}) are given by
\begin{align}
&{\cal A}_{11}(\omega) = \frac{1}{4}\Big[\delta(\omega + {\cal K} )+\delta(\omega - {\cal K} )+\delta(\omega + {\cal L} )+\delta(\omega - {\cal L} )\Big]\nonumber\\
&-\frac{\alpha^2}{4\sqrt{h_z^2\textbf{k}^2+\textbf{k}^2\alpha^2+\alpha^4}}\Big[\delta(\omega + {\cal K} )+\delta(\omega - {\cal K} )-\delta(\omega + {\cal L} )\nonumber\\
&-\delta(\omega - {\cal L} )\Big] + \frac{h_z}{2({\cal K}^2-{\cal L}^2)}\Big[ {\cal K}\{\delta(\omega + {\cal K} )-\delta(\omega - {\cal K} )\} \nonumber\\
&- {\cal L}\{\delta(\omega + {\cal L} )-\delta(\omega - {\cal L})\}\Big] - \frac{h_z^3-h\textbf{k}^2+4h\alpha^2}{2{\cal K}{\cal L}({\cal K}^2-{\cal L}^2)}\Big[ \nonumber\\
&{\cal L}\{\delta(\omega + {\cal K} )-\delta(\omega - {\cal K} )\} - {\cal K}\{\delta(\omega + {\cal L} )-\delta(\omega - {\cal L})\}\Big],
\end{align} 

\begin{align}
&{\cal A}_{22}(\omega) = \frac{1}{4}\Big[\delta(\omega + {\cal K} )+\delta(\omega - {\cal K} )+\delta(\omega + {\cal L} )+\delta(\omega - {\cal L} )\Big]\nonumber\\
&+\frac{\alpha^2}{4\sqrt{h_z^2\textbf{k}^2+\textbf{k}^2\alpha^2+\alpha^4}}\Big[\delta(\omega + {\cal K} )+\delta(\omega - {\cal K} )-\delta(\omega + {\cal L} )\nonumber\\
&-\delta(\omega - {\cal L} )\Big] - \frac{h_z}{2({\cal K}^2-{\cal L}^2)}\Big[ {\cal K}\{\delta(\omega + {\cal K} )-\delta(\omega - {\cal K} )\} \nonumber\\
&- {\cal L}\{\delta(\omega + {\cal L} )-\delta(\omega - {\cal L})\}\Big] + \frac{h_z^3-h\textbf{k}^2}{2{\cal K}{\cal L}({\cal K}^2-{\cal L}^2)}\Big[ \nonumber\\
&{\cal L}\{\delta(\omega + {\cal K} )-\delta(\omega - {\cal K} )\} - {\cal K}\{\delta(\omega + {\cal L} )-\delta(\omega - {\cal L})\}\Big],
\end{align} 

\begin{align}
&{\cal A}_{33}(\omega) = \frac{1}{4}\Big[\delta(\omega + {\cal K} )+\delta(\omega - {\cal K} )+\delta(\omega + {\cal L} )+\delta(\omega - {\cal L} )\Big]\nonumber\\
&+\frac{\alpha^2}{4\sqrt{h_z^2\textbf{k}^2+\textbf{k}^2\alpha^2+\alpha^4}}\Big[\delta(\omega + {\cal K} )+\delta(\omega - {\cal K} )-\delta(\omega + {\cal L} )\nonumber\\
&-\delta(\omega - {\cal L} )\Big] + \frac{h_z}{2({\cal K}^2-{\cal L}^2)}\Big[ {\cal K}\{\delta(\omega + {\cal K} )-\delta(\omega - {\cal K} )\} \nonumber\\
&- {\cal L}\{\delta(\omega + {\cal L} )-\delta(\omega - {\cal L})\}\Big] - \frac{h_z^3-h\textbf{k}^2}{2{\cal K}{\cal L}({\cal K}^2-{\cal L}^2)}\Big[ \nonumber\\
&{\cal L}\{\delta(\omega + {\cal K} )-\delta(\omega - {\cal K} )\} - {\cal K}\{\delta(\omega + {\cal L} )-\delta(\omega - {\cal L})\}\Big],
\end{align}

\begin{align}
&{\cal A}_{44}(\omega) = \frac{1}{4}\Big[\delta(\omega + {\cal K} )+\delta(\omega - {\cal K} )+\delta(\omega + {\cal L} )+\delta(\omega - {\cal L} )\Big]\nonumber\\
&-\frac{\alpha^2}{4\sqrt{h_z^2\textbf{k}^2+\textbf{k}^2\alpha^2+\alpha^4}}\Big[\delta(\omega + {\cal K} )+\delta(\omega - {\cal K} )-\delta(\omega + {\cal L} )\nonumber\\
&-\delta(\omega - {\cal L} )\Big] - \frac{h_z}{2({\cal K}^2-{\cal L}^2)}\Big[ {\cal K}\{\delta(\omega + {\cal K} )-\delta(\omega - {\cal K} )\} \nonumber\\
&- {\cal L}\{\delta(\omega + {\cal L} )-\delta(\omega - {\cal L})\}\Big] + \frac{h_z^3-h\textbf{k}^2+4h\alpha^2}{2{\cal K}{\cal L}({\cal K}^2-{\cal L}^2)}\Big[ \nonumber\\
&{\cal L}\{\delta(\omega + {\cal K} )-\delta(\omega - {\cal K} )\} - {\cal K}\{\delta(\omega + {\cal L} )-\delta(\omega - {\cal L})\}\Big],
\end{align} 

\begin{align}
&{\cal A}_{21}(\omega) = \frac{ih_z(k_x+ik_y)}{{\cal K}{\cal L}({\cal K}^2-{\cal L}^2)}\Big[ {\cal L}\{\delta(\omega + {\cal K} )-\delta(\omega - {\cal K} )\} \nonumber\\
&- {\cal K}\{\delta(\omega + {\cal L} )-\delta(\omega - {\cal L})\}\Big] + \frac{i(k_x+ik_y)}{4\sqrt{h_z^2\textbf{k}^2+\textbf{k}^2\alpha^2+\alpha^4}}\Big[ \nonumber\\
&\delta(\omega + {\cal K} )+\delta(\omega - {\cal K} ) -\delta(\omega + {\cal L} )-\delta(\omega - {\cal L})\Big],
\end{align} 

\begin{align}
&{\cal A}_{34}(\omega) = \frac{ih_z\alpha (k_x-ik_y)}{{\cal K}{\cal L}({\cal K}^2-{\cal L}^2)}\Big[ {\cal L}\{\delta(\omega + {\cal K} )-\delta(\omega - {\cal K} )\} \nonumber\\
&- {\cal K}\{\delta(\omega + {\cal L} )-\delta(\omega - {\cal L})\}\Big] - \frac{i\alpha(k_x-ik_y)}{4\sqrt{h_z^2\textbf{k}^2+\textbf{k}^2\alpha^2+\alpha^4}}\Big[ \nonumber\\
&\delta(\omega + {\cal K} )+\delta(\omega - {\cal K} ) -\delta(\omega + {\cal L} )-\delta(\omega - {\cal L})\Big],
\end{align} 
\begin{align}
{\cal A}_{12}(\omega)={\cal A}_{21}^*(\omega),
\end{align}
\begin{align}
{\cal A}_{43}(\omega)={\cal A}_{34}^*(\omega),
\end{align}
\begin{align}
{\cal K} = \Big[ h_z^2 + \textbf{k}^2 + 2\alpha^2 +2\sqrt{h_z^2\textbf{k}^2+\textbf{k}^2\alpha^2 +\alpha^4}\Big]^\frac{1}{2},
\end{align}
\begin{align}
{\cal L} = \Big[ h_z^2 + \textbf{k}^2 + 2\alpha^2 -2\sqrt{h_z^2\textbf{k}^2+\textbf{k}^2\alpha^2 +\alpha^4}\Big]^\frac{1}{2}.
\end{align}
The Dirac delta function is denoted by $\delta(X)$. Here we have considered $\textbf{h}=(0,0,h_z)$ to simplify the expressions. 
When all three components of 
the magnetization are nonzero and SOC is present, the resultant expressions 
become very lengthy, and
thus they are not presented here.

If we set spin-orbit coupling to zero, i.e., $\alpha=0$, and all three components of 
the Zeeman field can be nonzero, i.e., $\textbf{h}=(h_x,h_y,h_z)$, the Green's functions reduce to
\begin{subequations}
\begin{align}\label{g11_app}
G_{11}(\omega,\textbf{k},\textbf{h})=\frac{1}{2}\Big\{ \frac{1}{i\omega-\textbf{h}+\textbf{k}} + \frac{1}{i\omega-\textbf{h}-\textbf{k}}\Big\},
\end{align} 
\begin{align}
G_{22}(\omega,\textbf{k},\textbf{h})=\frac{1}{2}\Big\{ \frac{1}{i\omega+\textbf{h}-\textbf{k}} + \frac{1}{i\omega+\textbf{h}+\textbf{k}}\Big\},
\end{align} 
\begin{align}
G_{33}=G_{11},
\end{align} 
\begin{align}
G_{44}=G_{22}.\label{g44_app}
\end{align} 
\end{subequations}
Substituting these
 Green's functions (\ref{g11_app})-(\ref{g44_app}) into Eq.~(\ref{sig_ab}), we find,
\begin{align}
&\sigma_{xx}(\omega,\mu)=\frac{2\pi^2e^2}{\hbar}\int \frac{d\textbf{k}}{(2\pi)^2}\int \frac{d\varepsilon_n}{2\pi} {\cal F}(\varepsilon_n,\omega,\mu)\times \nonumber \\
&\Big\{ \Big[\delta(\varepsilon_n^+ +\textbf{k}^-)+\delta(\varepsilon_n^+-\textbf{k}^+)\Big]  \Big[\delta(\varepsilon_n+\textbf{k}^-)+\delta(\varepsilon_n-\textbf{k}^+)\Big]+\nonumber\\
&\Big[\delta(\varepsilon_n^+ -\textbf{k}^-)+\delta(\varepsilon_n^++\textbf{k}^+)\Big]  \Big[\delta(\varepsilon_n-\textbf{k}^-)+\delta(\varepsilon_n+\textbf{k}^+)\Big]
\Big\}.
\end{align} 
in which $\textbf{k}^{\pm}=\textbf{k}\pm \textbf{h}$. Note that in the absence of SOC, 
the off-diagonal Green's functions play no role in the optical conductivity 
and the transverse Hall components $\sigma_{xy, yx}(\omega)$ vanish altogether.

\section{Maxwell's equation, EM fields, and boundary conditions}\label{appx:MAXWELL}
Assuming a harmonic time dependence $\exp(-i\omega t)$ for the EM field, 
we have:
\begin{subequations}\label{d}
\begin{align} 
{\bm \nabla} \times {\bm E}_n &= i \omega \mu_0 \bar{\bar\mu}_n {\cdot} {\bm H}_n, \\
{\bm  \nabla} \times {\bm H}_n &= -i\omega \varepsilon_0 \bar{\bar \varepsilon}_n { \cdot} {\bm E}_n,
  \label{d1}
\end{align}
\end{subequations}
where the integer $n=0,1,2,3$ denotes the region (0 for the vacuum, 1 for graphene, 2 for the insulator layer, 
and 3 for the metallic region). Combining Eqs.~(\ref{d}), we obtain,
\begin{subequations}
\begin{align} 
\label{Maxw}
{\bm \nabla}\times\bigl( \bar{\bar\mu}_n^{-1} \cdot {\bm \nabla}\times{\bm E}_n\bigr)  &=
\, k_0^2\bigl(\bar{\bar\varepsilon}_n \cdot{\bm E}_n\bigr), \\
{\bm \nabla} \times\bigl( \bar{\bar\varepsilon}_n^{-1} \cdot {\bm \nabla}\times{\bm H}_n\bigr)  &=
\, k_0^2\bigl(\bar{\bar\mu}_n \cdot{\bm H}_n\bigr),
\end{align}
\end{subequations}
where the free-space wave number is $k_0=\omega/c$. 
For the nonmagnetic regions,
 we have
$\bar{\bar \mu}_n= \bar{\bar I}$. 
For the relative permittivity tensors, 
$\bar{\bar{\epsilon}}_0 =  \bar{\bar I}$ in the upper vacuum region, while 
 the  spacer layer is an isotropic dielectric with $\bar{\bar{\epsilon}}_2 = \epsilon_2 \bar{\bar I}$.
 The bottom region has $\bar{\bar{\epsilon}}_3 = \epsilon_3\bar{\bar I}$ (see Fig.~\ref{fig1}).

For a normally incident EM wave,
the electric   field is written,
${\bm E}_0 = {\bm E}_{i}+{\bm E}_{r}$, where
the incident and reflected fields are expressed as ${\bm E}_{i} = {\bm E}_{0i} e^{i k_{0} z}$,
and
${\bm E}_{r} ={\bm E}_{0r} e^{-i k_{0} z}$, respectively.
The 
   spin-orbit coupling and static magnetic field generate
  off-diagonal components to the permittivity tensor in 
  the graphene film.
The reflected electric field has components, ${\bm E}_{0r}=(E_{0rx},E_{0ry})$.
 Similarly, within the dielectric layer, 
the electric field is expressed as 
a superposition of
upward and downward propagating waves: 
${\bm E}_{2} ={\bm E}_{u}+{\bm E}_{d}$,
where 
${\bm E}_{u}={\bm E}_{2u}e^{-i k_{2} z}$,
 ${\bm E}_{d}={\bm E}_{2d}e^{i k_{2} z}$, and 
$k_2=k_0\sqrt{\epsilon_2}$.
The transmitted field in the  region below the spacer layer is 
${\bm E}_{t} = {\bm E}_{3t} e^{i k_{3} z}$,
where $k_3=k_0\sqrt{\epsilon_3}$ (we later take the limit of
a perfect metal in region 3). From Eq.~(\ref{d1}), the corresponding magnetic field 
in the vacuum region can be written ${\bm H}_0 = {\bm H}_{i}+{\bm H}_{r}$,
with
\begin{subequations}
\begin{align} 
&{\bm H}_{i} = -(E_{iy}/\eta_0)  \hat{  \bm x},\\
&{\bm H}_{r} = (1/\eta_0)(+E_{ry}\hat{  \bm x}- E_{rx} \hat{\bm y}),\\ 
&{\bm H}_{t} = (1/\eta_3)(-E_{ty}\hat{  \bm x} +E_{tx} \hat{\bm y}),
\end{align}
\end{subequations}
where ${\eta_0}=\sqrt{\mu_0/\epsilon_0}$ is the impedance of free space.
The magnetic field in the dielectric region can be decomposed as
${\bm H}_2={\bm H}_u+{\bm H}_d$, where
\begin{subequations}
\begin{align} 
&{\bm H}_{u} = (1/\eta_2)( +E_{uy}\hat{  \bm x} - E_{ux}\hat{\bm y}),\\
&{\bm H}_{d} = (1/\eta_2)(- E_{dy}\hat{  \bm x} + E_{dx} \hat{\bm y}).
\end{align}
\end{subequations}
Here
$\eta_2 = \eta_0/\sqrt{\epsilon_2}$ is the impedance of the dielectric layer,
and $\eta_3 = \eta_0/\sqrt{\epsilon_3}$ for the metal region.

The
presence of graphene enters in the boundary condition for the 
tangential components of the magnetic field by writing
\begin{align}
\hat{\bf n} \times ({\bm H}_0(\omega)-{\bm H}_2(\omega))=  {\bm J}(\omega),
\end{align}
where $\hat{\bf n}$ is the normal to the vacuum/graphene interface, and ${\bm J}$ is the
current density at the interface.
Thus, we have 
\begin{align} 
( H_{0y}-H_{2y} )|_{z=0} &=   (\sigma_{xx} E_{0x} + \sigma_{xy} E_{0y})|_{z=0}, \\
(H_{2x} - H_{0x})|_{z=0} &=  ( \sigma_{yx} E_{0x} + \sigma_{yy} E_{0y})|_{z=0}, 
\end{align}
where we used Ohm's law to connect 
the surface current density $(J_x(\omega),J_y(\omega))$  to the electric field: 
${\bm J} (\omega){=} \bar {\bar \sigma}{\bm E}(\omega)$. 
Note that one can also consider the graphene layer as a finite-sized slab,
like the spacer layer,
and using the dielectric response tensor discussed earlier,
solve for the fields within the graphene layer 
(for continuity, the calculations are
not shown here).
We have found that  this approach leads to equivalent results, but 
treating the graphene layer as a current sheet with infinitesimal  
thickness
leads to simpler expressions. 
Hence, we follow the latter approach.
We also have for the electric field at the graphene interface
\begin{align}
\hat{\bf n} \times ({\bm E}_0(\omega)-{\bm E}_2(\omega))=  0.
\end{align}
Upon matching the tangential ${\bm E}(\omega)$  fields at the vacuum/graphene
and dielectric/substrate interfaces, 
we have the following conditions:
\begin{align} \label{Econd}
&{\bm E}_{0i}  + {\bm E}_{0r}  - {\bm E}_{2u} - {\bm E}_{2d} = 0,\\
&{\bm E}_{2u} e^{-i k_2 d}+ {\bm E}_{2d} e^{i k_2 d} - {\bm E}_{3t} e^{i k_3 d} = 0, 
\end{align}
while matching the tangential ${\bm H}(\omega)$ fields at the dielectric/substrate interface gives
\begin{align} \label{Hcond}
\frac{{\bm E}_{3t} e^{i k_3 d} }{\eta_3} + \frac{{\bm E}_{2u} e^{-i k_2 d} }{\eta_2}
- \frac{{\bm E}_{2d} e^{i k_2 d} }{\eta_2}  = 0.
\end{align}

\section{Calculation of electromagnetic field coefficients}\label{appx:EMWs}

In the limit of a perfectly conducting substrate,
the electric field in the spacer region can be written simply as
${\bm E}_{2} =( E_{2x},E_{2y}) \sin(k_2(z-d))$.
The coefficients for the electromagnetic fields in the spacer and
vacuum regions are found by implementation of the appropriate boundary conditions discussed above. We
subsequently  find
the respective $x$ components of the reflection coefficient, and electric field component 
in the spacer layer to be
\begin{widetext}
\begin{align} \label{reflect1}
E^\pm_{0rx}& =\frac{
\left[ \eta_0 (\pm 2 i \sigma_{xy} - \eta_0 \sigma_{xy} \sigma_{yx} +  \eta_0 \sigma_{\parallel}^2)-1\right] \sin^2(k_2 d) 
+ i \kappa_2\eta_0  \sigma_{\parallel} \sin(2 k_2 d)  -\kappa_2^2 \cos^2(k_2 d) }
{
\kappa_2^2 \cos^2( k_2 d)
+\left [ \eta_0 (\eta_0 \sigma_{xy} \sigma_{yx} - 2 \sigma_{\parallel} -  \eta_0 \sigma_{\parallel}^2)-1\right]\sin^2(k_2 d) - 
 i \kappa_2 (1 + \eta_0 \sigma_{\parallel}) \sin(2 k_2 d)
 }, 
 \end{align}
\end{widetext}
and,
\begin{widetext}
\begin{align}
E_{2x}^\pm &=\frac{2 \left[1 + \eta_0 (\sigma_{\parallel}\mp i \sigma_{xy} ) + i \kappa_2 \cot(k_2d)\right]\csc(k_2 d)}
{
 \eta_0 (\eta_0 \sigma_{xy} \sigma_{yx} - 2 \sigma_{\parallel} - \eta_0 \sigma_{\parallel}^2)-1 + \kappa_2 \cot(k_2 d) \left[  \kappa_2 \cot(k_2 d)-2 i (1 + \eta_0 \sigma_{\parallel})\right]
}.
\end{align}
\end{widetext}
Here we have defined $\sigma_{\parallel}\equiv \sigma_{xx} = \sigma_{yy} $.
The reflection coefficient for the incident electric field in the $y$ direction
is found through the relationship
\begin{widetext}
\begin{align} \label{reflect2}
\frac{ E^\pm_{0ry}}{E^\pm_{0rx}}=\pm i -\frac{ 2 \eta_0 (\sigma_{xy} + \sigma_{yx}) \sin^2(k_2 d)}
{
\kappa_2^2 \cos^2( k_2 d ) + 
\left[1 + \eta_0 ( \eta_0 \sigma_{xy} \sigma_{yx} \mp 2 i \sigma_{xy}-\eta_0 \sigma_{\parallel}^2)\right] \sin^2(k_2 d) 
- i \kappa_2 \eta_0  \sigma_{\parallel} \sin(2 k_2 d)
}.
\end{align}
\end{widetext}
Similarly, the electric field coefficients for the spacer region are related via
\begin{align}
\frac{E_{2x}^\pm}{E_{2y}^\pm} &= \pm i- \frac{ \eta_0 (\sigma_{xy} + \sigma_{yx})}
{
1  +
   \eta_0 (\sigma_{\parallel}
   \mp
i  \sigma_{xy})
    + i \kappa_2 \cot(k_2 d) 
  }.
  \end{align}
As the expressions above show, the crucial difference in the coefficients for the
RH and LH polarizations is the $i \sigma_{xy}$ term arising from the SOC and Zeeman field induced
in the graphene layer.

\end{document}